\documentclass[12pt,fleqn]{article}

\usepackage{amsmath, amssymb,graphicx}

\begin{document}

%%%%%%%%%%%%%%%%   TITLE    %%%%%%%%%%%%%%%%%%%%

\thispagestyle{empty}
\renewcommand{\thefootnote}{\fnsymbol{footnote}}

{\hfill \parbox{4cm}{
        HU-EP-03/47 \\
        AEI-2003-069 \\
}}

\bigskip\bigskip

\begin{center} \noindent \Large \bf
Properties of Chiral Wilson Loops.
\end{center}

\bigskip\bigskip\bigskip

\centerline{ \normalsize \bf Z. Guralnik$^{a}$ and B. Kulik
$^{b}$\footnote[1]{\noindent \tt zack@physik.hu-berlin.de,
bogdan.kulik@aei.mpg.de} }

\bigskip
\bigskip\bigskip

\centerline{$^a$ \it Institut f\"ur Physik} \centerline{\it
Humboldt-Universit\"at zu Berlin} \centerline{\it Newtonstra{\ss}e
15} \centerline{\it 12489 Berlin, Germany}
\bigskip
\centerline{$^b$ \it Max-Planck-Institut f\"ur Gravitationsphysik}
\centerline{\it Albert-Einstein-Institut} \centerline{\it Am
M\"uhlenberg 1, D-14476 Golm, Germany}
\bigskip\bigskip

\bigskip\bigskip

\renewcommand{\thefootnote}{\arabic{footnote}}

\centerline{\bf \small Abstract}
\medskip

{\small We study a class of Wilson Loops in ${\mathcal N} =4, D=4$
Yang-Mills theory belonging to the chiral ring of a ${\cal N}
=2,d=1$ subalgebra. We show that the expectation value of these
loops is independent of their shape.  Using properties of the
chiral ring,  we also show that the expectation value is
identically $1$. We find the same result for chiral loops in
maximally supersymmetric Yang-Mills theory in three, five and six
dimensions. In seven dimensions, a generalized Konishi anomaly
gives an equation for chiral loops which closely resembles the
loop equations of the three dimensional Chern-Simons theory.}

\newpage

\section{Introduction}

For a variety of reasons,  it may be useful to write the action of
a supersymmetric theory with D-dimensional \footnote{We will use
'D' for dimensions of the gauge theories and 'd' for dimensions of
the superspace they are represented in.} Lorentz invariance in
terms of a lower dimensional superspace.    This procedure was
developed originally in \cite{Marcus,Arkani-Hamed} and has been
helpful in extra-dimensional model building  and in studying field
theories of intersecting branes
\cite{Erdmenger:2002ex,Constable:2002xt,Constable:2002vt,SLAG}.
Another application is the derivation of auxiliary k-dimensional
bosonic matrix models which capture the holomorphic data of
supersymmetric theories in $4+k$ dimensions
\cite{Dijkgraaf:2003xk,Bena}. There is yet another possible use as
a tool for obtaining non-renormalization theorems,  but this has
yet to be explored in much detail. The use of a lower dimensional
superspace to obtain non- renormalization theorems was first
hinted at in \cite{Erdmenger:2002ex}.  There it was used to
suggest an alternative argument for the non-renormalization of the
metric on the Higg's branch of four-dimensional ${\cal N} =2$
gauge theories.  This argument rests on the fact that some of the
hypermultiplet kinetic terms arise from the superpotential in a
lower dimensional superspace.  In general,  D-terms in theories
with extended supersymmetry may become F-terms upon using a lower
dimensional superspace. Furthermore gauge connections in
directions transverse to the lower dimensional superspace may
become components of chiral superfields.  Thus holomorphic
constraints may apply to quantities which one might not have
expected.

In the subsequent discussion, we will use a lower dimensional
superspace to obtain a non-renormalization theorem for BPS Wilson
loops in maximally supersymmetric Yang-Mills theory in various
dimensions. In particular, we will study the four dimensional
${\cal N} =4$ gauge theory using a ${\cal N} =2, d=1$ superspace.
In this language, the superpotential is obtained from a
Chern-Simons action by replacing gauge fields with chiral
superfields. The diffeomorphism invariance of this superpotential
leads to strong constraints on the chiral ring, defined with
respect to ${\cal N} =2, d=1$ supersymmetry.   We will focus on
the equations satisfied by a class of Wilson loops belonging to
the chiral ring.

Wilson loops in ${\mathcal N} =4$ gauge theory may involve the
adjoint scalars and fermions as well as the gauge connections.
Such Wilson loops have received considerable attention, beginning
with \cite{Maldacena:1998im}.
%In Minkowski space the Wilson loop which
%is usually considered has the general form
%\begin{align}\label{minkloop}
%W_{\rm Minkowski} = {\rm tr} P \exp  i \int ds \left( A_{\mu}
%\frac{dx^{\mu}(s)}{ds} + X^m \frac{dy^m (s)}{ds} \right)
%\end{align}Upon a careful continuation to Euclidean
%space \cite{Drukker:1999zq},  one should consider
In Euclidean space, the Wilson loop which is usually considered has
the general form:
\begin{align}\label{eucloop}
W_{\rm Euclidean} = {\rm tr} P \exp  i \int ds \left( A_{\mu}
\frac{dx^{\mu}(s)}{ds} + iX^m \frac{dy^m (s)}{ds} \right)\, ,
\end{align} where $m =4 \dots 9$.  The path $x^{\mu}(s)$ must be closed for
the
Wilson loop to be gauge invariant,  while no such constraint
applies to the path $y^m(s)$. A special class of loops satisfying
$\dot{y}^2 = \dot{x}^2$ arises naturally in the context of minimal
surfaces in $AdS$ \cite{Fischler:1998,Drukker:1999zq}. Loops in this class are
at least locally BPS, which facilitates computation of the
expectation value in some special cases such as the circular
Wilson loop with fixed orientation in the $y^m$ directions
\cite{Erickson:2000af,Drukker:2000rr,Bianchi:2002gz}. This
circular loop is annihilated by a linear combination of poincar\'e
and special supersymmetries \cite{Bianchi:2002gz}.

Wilson loops which are globally BPS with respect to the
Poincar\'{e} supersymmetries were studied by Zarembo
\cite{Zarembo:2002an}.  For these loops,  $dy^m = dx^{\mu}
M_{\mu}^m$, where the matrix satisfies:
\begin{align}
M_{\mu}^m M_{\nu}^m = \delta_{\mu\nu}\, . \end{align}  When these
Wilson loops are extended in $\mathbb{R}^2$, $\mathbb{R}^3$ or
$\mathbb{R}^4$ they are respectively $1/4, 1/8$ or $1/16$ BPS. The
Wilson loops which we will study have $y^{i+3}(s) = x^i(s)$ for
$i=1,2,3$ with $x^4(s) = {\rm const}$ and $y^{7,8,9}(s) = 0$.
These belong to the class of $1/4$ and $1/8$ BPS loops for which
\begin{align} M = \begin{pmatrix} 1 & 0 & 0 & 0 & 0 & 0\cr 0 & 1
& 0 & 0 & 0 & 0\cr  0 & 0 & 1 & 0 & 0 & 0\cr 0&0&0&1&0&0
\end{pmatrix}\, .
\end{align}
It turns out that these Wilson loops belong to the chiral ring of
an ${\cal N} =2, d=1$ sub-algebra of the full ${\cal N}=4, D=4$
supersymmetry, and we will refer to them as ``chiral Wilson
loops''. Of course our results extend to any other loop related by
the Lorentz and R-symmetries,  which act in the obvious way on
$M_{\mu}^m$.

To the first two orders in the 't Hooft coupling and leading order
in the $1/N$ expansion, it has been shown that all contributions to the
expectation value of a globally BPS Wilson loop cancel
\cite{Zarembo:2002an}, giving expectation value $1$.
%The next
%order has also recently been computed \cite{AEIstudent}, and again
%gives no contribution.
At large 't Hooft coupling, AdS/CFT duality
may be used to compute the expectation value of a Wilson loop
\cite{Maldacena:1998im,Drukker:1999zq} by summing over minimal
surfaces in AdS which are bounded by the loop.  The minimal
surfaces associated with the circular and rectangular BPS loops
have zero regularized area \cite{Zarembo:2002an},  which together with a
counting of zero modes leads to the
expectation value $\exp(-A) =1$.  On this basis,  it was
conjectured \cite{Zarembo:2002an} that all $1/4$ BPS (planar)
Wilson loops are non-renormalized.

By considering the equations of motion of the ${\mathcal N}=4$
theory in ${\mathcal N} =2, d=1$ superspace, we will obtain a loop
equation showing that the expectation values of a chiral Wilson
loop is independent of its shape. Together with properties of the
chiral ring,  this can be used to show that the expectation value
is identically $1$.  This holds whether the loop is $1/4$ BPS
(planar) or $1/8$ BPS (extended in $\mathbb{R}^3$).  This result
is somewhat stronger than the conjecture of \cite{Zarembo:2002an},
which proposed non-renormalization of the $1/4$ BPS loops. The
reason a stronger proposal was not made in \cite{Zarembo:2002an}
is the following.   Besides the area of the minimal surface, the
AdS computation of the loop expectation value also depends on the
number of zero modes of the minimal surfaces. For loops in
$\mathbb{R}^3$, a counting of the most transparent zero modes
indicates a non-trivial dependence on the 't Hooft coupling even
if the regularized area is zero. However it is possible that there
are other less obvious zero modes which were
missed\footnote{Private conversation with K. Zarembo.}.

Our results concerning the non-renormalization of chiral Wilson
loops in ${\mathcal N}=4$ Yang-Mills are easily extended to
maximally supersymmetric Yang-Mills theory in dimensions $D= 3,5$
and $6$ by using a four supercharge $d=D-3$ dimensional
superspace. For $D=7$ however, we find a generalized Konishi
anomaly which ruins the shape independence of the loop expectation
value. The Konishi anomaly closely resembles the loop equations of
three dimensional Chern-Simons theory. For dimensions $D>7$, our
superspace formalism is no longer applicable.
%a superspace with
%$d>4$ is associated with more than four supercharges.

\section{$D$-dimensional maximally supersymmetric Yang- $\qquad$
    Mills in $D-3$ dimensional superspace.}

To obtain constraints on expectation values of certain BPS Wilson
loops, we will make use of a formalism in which these loops
manifestly belong to a chiral ring with respect to a lower
dimensional supersymmetry algebra. In order to include gauge
connections in chiral superfields, it is necessary to use a
superspace with dimension less than that of the theory which one
is studying. In this section, we will write the action of
maximally supersymmetric D-dimensional Yang-Mills theory in terms
of a four-supercharge $d=D-3$ dimensional superspace.

\subsection{${\cal N} =4$ SYM in quantum mechanical superspace}

We wish to write the four-dimensional ${\cal N} =4$ SYM action in
a one-dimensional ${\cal N} =2$ superspace.   This is easily done
by dimensional reduction of a result from the literature, in which
16-supercharge 7-dimensional Yang-Mills is written in terms of
${\cal N} =1, d=4$ superspace \cite{Arkani-Hamed}\footnote{A very
similar procedure was discussed earlier in \cite{Marcus}, where
16-supercharge ten-dimensional Yang-Mills was written in ${\cal
N}=1, d=4$ superspace.}. This superspace representation of the
${\mathcal N} =4$ theory was recently discussed in \cite{SLAG},
where a relation between F and D-flatness and the calibration
equations for a special Lagrangian manifold was demonstrated.

The one-dimensional ${\cal N}=2$ superspace resembles the familiar
four dimensional ${\cal N} = 1$ superspace, and can be obtained
from it by dimensional reduction on $T^3$.  We emphasize however
that we are not dimensionally reducing the ${\cal N}=4$ theory,
but writing the full four-dimensional action in terms of a
one-dimensional superspace.

The ${\cal N} =2, d=1$ superfields entering the action for the
four-dimensional ${\cal N} =4$ Yang-Mills theory have the general
form $F(t,\theta,\bar\theta | \vec x)$, where
$(t,\theta,\bar\theta)$ spans the one dimensional superspace,  and
$\vec x \sim (x^1,x^2,x^3)$ can be regarded as continuous indices.
The necessary degrees of freedom are contained in three chiral
fields $\Phi_i$ and a vector field $V$. The vector superfield
satisfies $V= V^{\dagger}$.  Chiral superfields $\Phi$ satisfy
$\bar D_{\alpha} \Phi =0$, where:
\begin{align}
\bar D_{\alpha} =  - \frac{\partial}{\partial \bar
\theta^{\alpha}} -i\theta^{\alpha} \partial_t\,\qquad D_{\alpha} =
- \frac{\partial}{\partial \theta^{\alpha}} -i \bar\theta^{\alpha}
\partial_t\, .
\end{align}
The index $\alpha =1,2$ is associated with the $SU(2)$ R-symmetry
of ${\mathcal N}=2, d=1$.  The action of the four dimensional
${\cal N} =4$ theory in ${\cal N} =2, d=1$ superspace is:
\begin{align}\label{act}
S = & \frac{1}{g^2} \int d^3x\, dt\, \left. {\rm tr}\, \left[
{\cal W}_{\alpha}{\cal W}^{\alpha} + \epsilon_{ijk}(\Phi_i
\frac{\partial}{\partial x^j} \Phi_k +
\frac{2}{3}i\Phi_i\Phi_j\Phi_k) \right]\right|_{\theta\theta} +
cc\,  + \nonumber\\ & \frac{1}{g^2} \left. \int d^3x\, dt\, {\rm
tr}\, \bar\Omega_i e^V \Omega_i e^{-V}
\right|_{\theta\theta\bar\theta\bar\theta}\, ,
\end{align}
where the indices $i,j,k$ take values from $1$ to $3$, $W_{\alpha}
= \bar D\bar D e^V D_{\alpha} e^{-V}$, and:
\begin{align} \Omega_i \equiv \Phi_i + e^{-V}(i\partial_i -
\bar\Phi_i)e^{V}\, . \end{align}

Although the superfield content of the ${\cal N} =4$ theory in
${\cal N}=2,d=1$ superspace is similar to that in ${\cal
N}=1,d=4$ superspace, the component fields are distributed very
differently.  The bosonic fields of the ${\mathcal N} =4$ theory
consist of four components of the gauge connections $A_{0,1,2,3}$
and six Hermitian adjoint scalars $X_{4,5,6,7,8,9}$. These are
distributed amongst the ${\mathcal N} =2, d=1$ superfields $V$ and
$\Phi_i$ as follows:
\begin{align}
V \rightarrow A_0, X^{7,8,9}  \qquad \Phi_i \rightarrow A_i,
X^{i+3}\, .
\end{align}
The combination $A_i + iX^{i+3}$ is the bottom component of the
chiral super-field $\Phi_i$. Gauge transformations are
parameterized by ${\cal N} =2, d=1$ chiral superfields
$\Lambda(t,\theta; \vec x)$ which act in the following way:
\begin{align}
e^{V} &\rightarrow
e^{i\Lambda^{\dagger}}e^{V}e^{-i\Lambda}\\
\Phi_i &\rightarrow e^{i\Lambda} \Phi_i e^{-i\Lambda} -
e^{i\Lambda} i\frac{\partial}{\partial x^i}e^{-i\Lambda}\, ,
\end{align}
under which:
\begin{align}
\Omega_i \rightarrow e^{i\Lambda} \Omega_i e^{-i\Lambda}\, .
\end{align}
Four dimensional Lorentz invariance is not manifest, but becomes
apparent upon integrating out F and D-terms.

Note that the superpotential resembles a Chern-Simons action and
is diffeomorphism invariant,  although the theory as a whole is
not. We shall take advantage of this feature to obtain information
about a class of Wilson Loops which are chiral with respect to the
one dimensional ${\cal N} =2$ supersymmetry algebra.

\subsection{Euclidean action}

One can also write the Euclidean ${\mathcal N} =4, D=4$ action in
one dimensional superspace.  To this end, we start with the
Minkowski-space action for a D6-brane (7-dimensional maximally
supersymmetric Yang-Mills), which in four-dimensional ${\mathcal
N} =1$ superspace is:
\begin{align}\label{d6brane}
S = & \frac{1}{g^2} \int d^3x\, d^4y\, \left. {\rm tr}\, \left[
{\cal W}_{\alpha}{\cal W}^{\alpha} + \epsilon_{ijk}(\Phi_i
\frac{\partial}{\partial x^j} \Phi_k +
\frac{2}{3}i\Phi_i\Phi_j\Phi_k) \right]\right|_{\theta\theta} +
cc. +\\ & \frac{1}{g^2} \left. \int d^3x\, d^4y\, {\rm tr}\,
\bar\Omega_i e^V \Omega_i e^{-V}
\right|_{\theta\theta\bar\theta\bar\theta}\, .
\end{align}
Compactifying the time direction $y^0$ and two spatial directions
$y^1, y^2$ belonging to the four dimensional superspace gives an
action of the same form as (\ref{act}), the only difference coming
from the definition of the operators $D_{\alpha}$ and.$\bar
D_{\alpha}$.
%Alternatively,  one can make a more conventional analytic
%continuation from Minkowski to Euclidean space which does not
%preserve an extended supersymmetry
Note that the R-symmetry associated with the four-dimensional
Euclidean ${\mathcal N}=4$ supergroup is $Spin(1,5)$.  Non-compact
R-symmetry groups\footnote{Associated with the non-compact
R-symmetry group is a ``wrong sign'' kinetic term for the scalar
arising from $A_0$ upon compactification.  Thus it is not obvious
how to define the Euclidean theory non-perturbatively.  One
possibility is that, despite the wrong sign kinetic term, the
Schwinger-Dyson equations (which include the loop equations we
will later consider) have path integral solutions in which real
fields are extended to complex fields and one integrates over the
appropriate convergent contour in the complex plane.} are a well
known feature of Euclidean theories with extended supersymmetry.
For a discussion of Euclidean supersymmetry see
\cite{Zumino:1977yh,vanNieuwenhuizen:1996tv,vanNieuwenhuizen:1996ip,
Blau:1997pp,Belitsky:2000ii,Theis:2001ef}.

\subsection{Other dimensions}

The above discussion is readily generalized to maximally
supersymmetric Yang-Mills theories in $D=3,5,6$ and $7$
dimensions. In a four supercharge $d=D-3$ dimensional superspace,
the action of the theory is just:
\begin{align}\label{other}
S = & \frac{1}{g^2} \int d^3x\, d^dy\, \left. {\rm tr}\, \left[
{\cal W}_{\alpha}{\cal W}^{\alpha} + \epsilon_{ijk}(\Phi_i
\frac{\partial}{\partial x^j} \Phi_k +
\frac{2}{3}i\Phi_i\Phi_j\Phi_k) \right]\right|_{\theta\theta} +
cc. +\\ & \frac{1}{g^2} \left. \int d^3x\, d^dy\, {\rm tr}\,
\bar\Omega_i e^V\Omega_i e^{-V}
\right|_{\theta\theta\bar\theta\bar\theta}\, .
\end{align}
Note that for $d=0$,  the action can be viewed as that of a
supersymmetric matrix model, with an infinite number of matrix
superfields labelled by $\vec x$.  The supersymmetry generators
are:
\begin{align}
Q_{\alpha}= \frac{\partial}{\partial \theta^{\alpha}}, \qquad \bar
Q_{\alpha}= \frac{\partial}{\partial \bar\theta^{\alpha}}\, .
\end{align}
Chiral superfields $\Phi$ are defined by $\bar D_{\alpha} \Phi
=0$, where:
\begin{align}
D_{\alpha}= \frac{\partial}{\partial \theta^{\alpha}}, \qquad \bar
D_{\alpha}= \frac{\partial}{\partial \bar\theta^{\alpha}}\, ,
\end{align}
such that $\Phi = \phi + \theta \psi + \theta\theta F$.  Vector
superfields are defined as usual by $V = \bar V$.  The three gauge
connections and seven adjoint scalars of the theory are
distributed amongst these superfields as follows:
\begin{align}
V \rightarrow X^{7,8,9,10},  \qquad \Phi_i \rightarrow A_i,
X^{i+3}\, ,
\end{align}
where $i=1,2,3$.  As always, three of the gauge connections belong
to the lowest component of chiral superfields.  Upon integrating
out auxiliary fields, one finds the Euclidean action of the
maximally supersymmetric $D=3$ Yang-Mills theory \footnote{Note
that $V = \theta\bar \theta X^7 + \theta \sigma^1 \bar \theta X^8
+ \theta \sigma^2 \bar \theta X^9 + \theta \sigma^3 \bar \theta
X^{10} + \cdots$ so that $X^7$ is like a dimensional reduction of
a time-like gauge connection, and has a ``wrong sign'' kinetic
term. This is consistent with the R-symmetry of the Euclidean
supergroup, which is $SO(1,6)$.}.  It should be noted that for the
zero dimensional superspace there is no chiral ring in the usual
sense. The expectation value of a product of chiral operators is
not necessarily the product of the expectation values.  Since the
superspace is $0$ dimensional, one can not make the usual cluster
property argument to show that the expectation value of the
product is the product of the expectation values.

\section{Chiral Loop equations}

Consider the variation:
\begin{align}\label{trans}
\Phi_i(\vec x) \rightarrow \Phi_i(\vec x) + \epsilon g_{i, \vec
x}[\Phi]\, ,\end{align} where $\epsilon$ is a chiral function and
$g_{I,\vec x}$ is, for the moment, an arbitrary functional of
chiral superfields. The equations of motion which follow from this
variation are:
\begin{align}\label{Kon}
\bar D^2 \, {\rm tr}\, \left(g_{i,\vec x}[\Phi](e^{-V(\vec
x)}\bar\Omega_i(\vec x) e^{V(\vec x)} - \Omega_i(\vec x) ) \right)
+ {\rm tr}\, g_{i,\vec x}[\Phi]\frac{\delta W}{\delta \Phi_i(\vec
x)} = {\cal A}\, .
\end{align}
where ${\cal A}$ is a possible anomalous term which vanishes
classically.  Since $\Phi_i|_{\theta = \bar \theta =0}$ contains
the spatial gauge connections, let us choose $g_{i,\vec x} [\Phi]$
to be a spatial Wilson line on a contour ${\cal C}$ which begins
and ends at the point $\vec x$:
\begin{align}
g_{i,\vec x}[\Phi] = W(C,\vec x) = P \exp\left( i\int_{{\cal
C},\vec x}^{\vec x} \vec \Phi\cdot d\vec y\right)\, .
\end{align}
With this choice equation (\ref{Kon}) becomes:
\begin{align}\label{Konam}
\bar D^2( \cdots ) = {\rm tr} \left( W(C,\vec x) \epsilon_{ijk}
{\cal F}_{jk}(\vec x)\right) + {\cal A}\, ,
\end{align}
where:
\begin{align} {\cal F}_{jk} = \partial_j\Phi_k -
\partial_k\Phi_j + i[\Phi_j,\Phi_k]\, .
\end{align}
In the next section,  we shall argue that the anomaly vanishes, so
long as we are considering maximally supersymmetric Yang-Mills in
less than 7 dimensions.

The relation $(\bar D^2 J)|_{\theta=\bar\theta =0} = [\bar Q,
[\bar Q, J|_{\theta=\bar\theta=0}]$ holds for any gauge invariant
superfield $J$, such that $\langle\bar D^2 J|_{\theta=\bar\theta =0}\rangle
=0$ in a
supersymmetric vacuum. Therefore (\ref{Konam}) implies:
\begin{align}\label{expect}
\left< \left. {\rm tr} \left( W(C,\vec x) \epsilon_{ijk} {\cal
F}_{jk}(\vec x)\right) \right|_{\theta = \bar\theta =0} \right> =
\langle{\cal A}|_{\theta = \bar\theta =0}\rangle \, .
\end{align}
The insertion of the field strength ${\cal F}_{jk}$ generates an
infinitesimal deformation of the contour ${\cal C}$ in the $jk$
plane (see figure 1.).

\begin{figure}[!ht]
\begin{center}
\includegraphics[height=4cm,clip=true,keepaspectratio=true]{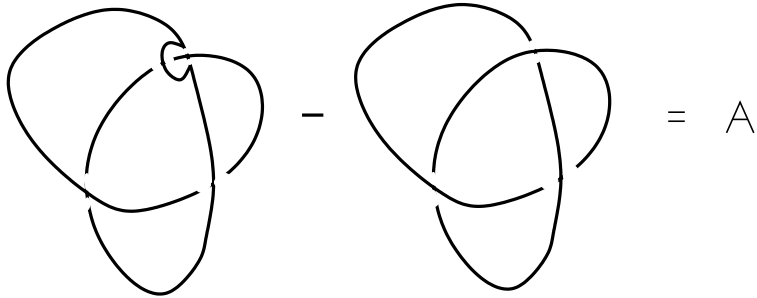}
\caption{(Almost) loop equation.}
\end{center}
\end{figure}

If the anomaly vanishes,  the chiral Wilson loops have no
dependence on their shape. This is a much stronger statement than
diffeomorphism invariance, since the expectation value does not even
have any dependence on the loop topology.  Note that there are no
terms in the loop equation which contribute when segments of the
loop cross each other.  There should be no ambiguities in
defining the expectation values of self intersecting Wilson loops.

Shape independence implies that the expectation value of the
chiral Wilson loop can only be a function of the gauge coupling
and theta angle in the conformal four dimensional case.  In
dimensions other than four,  shape independence implies that the
expectation value is a number.  This number can be shown to be $1$
by taking a  weak coupling limit, corresponding to shrinking the
loop in three dimensions and expanding the loop in five and more
dimensions.  In seven dimensions, we will argue that the anomaly
is non-zero.

In the conformal four dimensional case, shrinking the loop is ill
defined since there is no scale in the theory.  One might try to
show that the expectation value is still $1$ by using ``zig-zag''
symmetry. Zig-zag
symmetry means that segments of a loop which backtrack cancel each
other. This has been emphasized as a basic property of the QCD
string \cite{poly}, for which one only considers loops involving
the gauge field, but is not necessarily a property of Wilson loops
such as (\ref{eucloop}) which also involve the scalar fields. The
chiral Wilson loops satisfy zig-zag symmetry because of the
equivalence of the paths associated with the gauge field and the
scalars. Assuming that shape independence includes singular
deformations of the loop, zig-zag symmetry implies:
\begin{align}
\left<\frac{1}{N}{\rm tr} P \exp\left(i\oint_{\Sigma \in
\mathbb{R}^2} \phi\right)\right> = 1\, ,
\end{align}
through the manipulations illustrated in figure 2. Note however
that this requires the introduction of a cusp\footnote{To at least
the first two orders in perturbation theory,  there does not seem
to be anything special about a cusp, and the usual cancellations
still occur.}.
\begin{figure}[!ht]
\begin{center}
\includegraphics[height=2.5cm,clip=true,keepaspectratio=true]{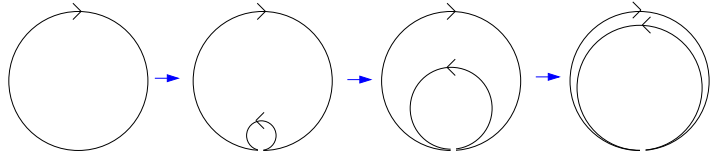}
\caption{Using zig-zag symmetry and shape independence to show
that $\langle\frac{1}{N}{\rm}trW(C)\rangle= 1$.  This assumes that
shape independence still holds when cusps are introduced.  An
argument which avoids this assumption relies on properties of the
chiral ring.}
\end{center}
\end{figure}

Fortunately, there is another argument that the expectation value
is one, which does not require passing through a loop with a cusp.
Given a Wilson-loop $\frac{1}{N}{\rm tr} W$ associated with a path
C in $\mathbb{R}^3$ one can smoothly deform the path within
$\mathbb{R}^3$ such that it goes around $C$ multiple times and the
Wilson loop becomes $\frac{1}{N}{\rm tr} W^n$ for any $n>1$. Shape
independence implies that the expectation value is unchanged:
\begin{align}\label{shp}
\langle\frac{1}{N}{\rm tr} W^n\rangle=\langle\frac{1}{N}{\rm tr}
W\rangle\, .  \end{align} Furthermore, there are relations amongst
the variables ${\rm tr} W^n$,  such that ${\rm tr} W^n$ for $n=1
\cdots 2N^2$ form a complete independent set\footnote{An example
of such a relation for the simpler case of $U(2)$ matrices $U$ is
$ 3({\rm tr} U)({\rm tr} U^2) - ({\rm tr} U)^3 = 2{\rm tr} U^3$.
The analogous relations for the matrices $W$ are more complicated
since these are generic $N\times N$ matrices with complex entries
and no constraints. The chiral Wilson loop is not the trace of a
unitary matrix because the exponent involves both hermitian and
anti-hermitian parts.}. Since the chiral Wilson loops belong to a
chiral ring, the expectation values factorize\footnote{This is
true provided that the superspace with respect to which the Wilson
loops are chiral is not zero dimensional.  For the three
dimensional ${\mathcal N} =8$ theory a different argument, such as
shrinking the Wilson loop, is required to show that
$\langle\frac{1}{N}{\rm tr} W\rangle =1$. See
\cite{Cachazo:2002ry} for a review of properties of the chiral
ring.}:
\begin{align}\label{factor}
\langle\frac{1}{N}{\rm tr} W^n \frac{1}{N}{\rm tr} W^m\rangle =
\langle\frac{1}{N}{\rm tr} W^n\rangle\langle \frac{1}{N}{\rm tr}
W^m\rangle \, .
\end{align}
The relations amongst the $\frac{1}{N}{\rm tr} W^n$, together with
(\ref{factor}) and (\ref{shp}) are solved by:
\begin{align}
\langle\frac{1}{N}{\rm tr} W^n\rangle = \frac{1}{N}{\rm
tr} {\cal W}_c\, ,
\end{align}
where ${\cal W}_c$ is a matrix (analogous to a master field)
satisfying ${\cal W}_c^2 = {\cal W}_c$. In the weak coupling limit
${\cal W}_c = I$. Assuming that the expectation value of the
Wilson loop depends smoothly on the coupling one has ${\cal W}_c =
I$ for any coupling, so that and $\langle\frac{1}{N}{\rm tr}
W\rangle = 1$.

Note that in the four dimensional ${\cal N} =4$ theory,  the
expectation value of the chiral Wilson loops is identically $1$ to
at least the first two orders in the 't Hooft coupling and leading
order in $1/N$ \cite{Zarembo:2002an}.  This is essentially due to
cancellations between greens functions of $A_i$ and $X^{i+3}$ due
to the relative factor of $i$ in the exponent of the Wilson loop,
$\phi_i = A_i + i X^{i+3}$.

\subsection{Variation of the functional measure.}

To see if there is a non-zero anomaly ${\mathcal A}$ in
(\ref{Konam}), we will compute the variation of the functional
measure under (\ref{trans}) using methods discussed in
\cite{Konishi:1985tu,Cachazo:2002ry}. The Jacobian is formally:
\begin{align}
J = \left|{\rm det}_c \frac{\delta
\Phi_i^{\prime}(Z^{\prime})}{\delta \Phi_j(Z)}\right| \, ,
\end{align}
where $Z$ collectively denotes the superspace coordinates $z \sim
\vec y, \theta,\bar\theta$ and the transverse spatial coordinates
$\vec x$. The subscript ``c'' in ${\rm det}_c$ stands for chiral
part of the whole super-determinant. It will be convenient to
define a Hilbert space spanned by $|Z,A\rangle$,  which is a
complete set of states in a superspace coordinate representation
satisfying:
\begin{align}
\hat Z |Z,A\rangle = Z |Z,A\rangle,\qquad  \hat T^A |Z,B\rangle =
f_{ABC} |Z,C\rangle \nonumber \\
\langle Z,A|Z^{\prime}, B\rangle\ = \delta_{AB}\delta^3(\vec x -
\vec x^{\prime}) \delta^d(y-y^{\prime}) \delta^4(\theta -
\theta^{\prime}),
\end{align}
where $\hat T^A$ are $U(N)$ generators and $f_{ABC}$ are the
$U(N)$ structure functions. We can then write:
\begin{align}
\frac{\delta \Phi_i^A(Z^{\prime})}{\delta \Phi_j^B(Z)} =
\delta_{ij}\delta_{AB}\delta^3(\vec x - \vec x^{\prime})
\delta^d(y-y^{\prime}) \delta^2(\theta - \theta^{\prime}) =
<Z,A|\bar D^2|Z',B> \, .
%\delta_{ij}\delta^3(\vec x - \vec x^{\prime})<z|\bar D^2|z'>
%(-\bar D^2/4) \delta^8(z)
\end{align}
For the infinitesimal variation (\ref{trans}), the Jacobian is:
\begin{align}\label{jac}
J = 1 + {\rm tr}_c\, \epsilon \bar D^2 \frac{\delta g_{i,\vec
x}}{\delta \Phi_i(\vec x)} = \int dZ_c\,\sum_A\, \epsilon(Z)
\langle Z, A| \bar D^2 \frac{\delta g^i[\Phi]}{\delta \Phi^i}|Z,
A\rangle \, ,
\end{align}
where $dZ_c$ is the chiral measure $d^dy\, d^2\theta\, d^3x$. The
matrix:
\begin{align} \langle Z,A|  \frac{\delta g_{i,\vec
x}[\Phi]} {\delta \Phi_i(\vec x')} (-\bar D^2/4) |Z' A>\, ,
\label{thematrix}\end{align} is proportional to $\delta^2(\theta
-\theta') = (\theta - \theta')^2$ and so has vanishing diagonal
entries. Thus naively $J = 1$. However this is not necessarily
true upon regularizing the trace.

To obtain the Jacobian for the transformation $\Phi^i\rightarrow
\Phi^i + \epsilon g^i[\Phi]$, we need to regularize the diagonal
elements of the matrix (\ref{thematrix}):
\begin{align}\label{diag}
{\cal M}_{\vec x, z, A} \equiv\, <\vec x, z, A| \bar D^2
\frac{\delta g^i[\Phi]}{\delta \Phi^i}|\vec x, z, A> \, .
\end{align}
In the more familiar context of the Konishi anomaly in ${\mathcal
N}=1$ four dimensional gauge theory, this is accomplished
\cite{Konishi:1985tu} by the insertion of an operator $\exp(-\hat
L/M^2)$ where:
\begin{align}
\hat L \equiv -\frac{1}{16}\bar D^2 e^{-V}D^2e^V\, .
\end{align}
Note that this operator is gauge covariant and chiral.  However,
the insertion of $\exp(-\hat L/M^2)$ will not suffice in our case,
since this only cuts off large momenta in directions belonging to
the superspace, which is lower-dimensional in our case. The
regularized version of (\ref{diag}) which we will consider is:
\begin{align}\label{reg}
{\cal M}_{\vec x, z, A} \equiv\, <\vec x, z, A| \exp(-\hat{\cal
L}/M^2)\,\bar D^2 \, \frac{\delta g^i[\Phi]}{\delta \Phi^i}|\vec
x, z, A>\, ,
\end{align}
where: \begin{align} \hat {\cal L}
%\equiv -\frac{1}{16}\bar D^2
%e^{-V}D^2e^V
= \hat L + (\partial_i + \Phi_i)(\partial_i + \Phi_i)\, .
\end{align}
To evaluate (\ref{reg}), note that:
\begin{align}\label{hatl}
\hat L (\bar D^2\cdots) = (\partial_t^2 - {1/2}{\cal
W}^{\alpha}D_{\alpha} + C\partial_t + F)(\bar D^2 \cdots)\, ,
\end{align}
where:
\begin{align}
C \equiv \frac{1}{2}\bar D_{\alpha}e^{-V} \bar D_{\alpha} e^V
\nonumber \\
F \equiv \frac{1}{16}\bar D^2 e^{-V}D^2e^V \, .
\end{align}
We can write:
\begin{align}
e^{-{\cal L}/M^2} = e^{-\nabla/M^2} \hat {\cal S}\, ,
\end{align} where $\nabla$ is the
Laplacian in the space including all bosonic coordinates, $\nabla
\equiv \nabla_{\vec x}^2 + \nabla_{\vec y}^2$\, . The factor
$\hat{\cal S} = 1 + \dots$ must contain a term with two
$D_{\alpha}$ operators for $\exp(\hat{\cal L}/M^2)\bar D^2$ to
give a non-zero contribution to (\ref{reg}). To illustrate this
property, note that:
\begin{align} <z^{\prime}| D^2 \bar D^2|z> &= \delta^d(y-y') D^2
\bar D^2 \delta^2(\theta -\theta')\delta^2(\bar\theta -
\bar\theta') \nonumber \\ &=  \delta^d(y-y') D^2 \bar D^2
(\theta-\theta^{\prime})^2(\bar\theta - \bar\theta^{\prime})^2 =
\delta^d(y-y')\, .
\end{align}
If the $D^2$ were removed, the diagonal matrix element would
vanish.  Thus,  (\ref{hatl}) implies that, in a large $M^2$
expansion, the leading non-zero contribution to (\ref{reg}) is:
\begin{align}\label{expr1}
&{\cal M}_{\vec x, z, A} = \nonumber \\
&<\vec x, z, A|\exp(-\nabla/M^2)
\frac{W_{\alpha}W^{\alpha}}{M^4}D^2\bar D^2|\vec x', z', A'><\vec
x', z', A'|\frac{\delta g^i}{\delta \Phi^i} |\vec x, z, A> +
\cdots\, ,
\end{align}
where the summation over primed variables is assumed and:
\begin{align}
<\vec x', z', A'|\frac{\delta g^i}{\delta \Phi^i} |\vec x, z, A>
\equiv\, \frac{\delta g^i_{\vec x', A'}}{\delta \Phi^i_{\vec x,
A}} \delta^d(\vec y' -\vec y)\delta^2(\theta
-\theta')\delta^2(\bar\theta' -\bar\theta)\, .
\end{align}
Expression (\ref{expr1}) can be evaluated by inserting the
identity $|k_x,k_y,C><k_x,k_y,C|$\,\, after $\exp(-\nabla/M^2)$,
where $|k_x,k_y,C>$ is an eigenvector of the momentum operators in
the transverse space $\vec x$ and the bosonic part of the
superspace $z$. The result is:
\begin{align}
{\cal M}_{\vec x, z,A} =
\qquad\qquad\qquad\qquad\qquad\qquad\qquad
\qquad\qquad\qquad\qquad\qquad\nonumber \\
\int d^3k_x\, d^d k_y\, d^3x'\, d^dy'\, \exp\left(-\frac{\vec
k_x^2 + \vec k_y^2}{M^2} + i \vec k_x \cdot (\vec x-\vec x') +
i\vec k_y
\cdot(\vec y -\vec y')\right) \nonumber \\
\frac{1}{M^4}W_{\alpha}^D(\vec x', z') W^{\alpha E}(\vec x',z')
<A|\hat T^D \hat T^E| A'> \frac{\delta g^i_{\vec x', A'}}{\delta
\Phi^i_{\vec x, A}} \delta^d(\vec y' -\vec y) + \cdots  \nonumber \\
= M^{d-4}W_{\alpha}^D(\vec x, z) W^{\alpha E}(\vec x,z) <A|\hat
T^D \hat T^E| A'> \frac{\delta g^i_{\vec x, A'}}{\delta
\Phi^i_{\vec
x, A}} +\cdots \nonumber \\
= M^{d-4} \sum_{I,J}%i\frac{g^2}{32\pi^2}
\left[{\cal W}^{\alpha}(\vec x),\left[{\cal W}_{\alpha}(\vec x),
  \frac{\delta g_{i,\vec x}[\Phi]}{\delta {\Phi_i(\vec x)}_{IJ}}
\right]\right]_{IJ} + \cdots \, ,
\end{align}
where ``$\cdots$'' indicates sub-leading terms in $1/M$. The
indices $I$ and $J$ in the last line are $N\times N$ matrix
indices. After sending the regularization mass $M$ to infinity the
anomaly vanishes for $d<4$, corresponding to the maximally
supersymmetric Yang-Mills theory in dimensions $D<7$. For $D=7$
one has:
\begin{align}\label{sevenanom}
\left< \left. {\rm tr} \left( W(C,\vec x) \epsilon_{ijk} {\cal
F}_{jk}(\vec x)\right) \right|_{\theta = \bar\theta =0} \right> =
\left< \left.\sum_{I,J}%i\frac{g^2}{32\pi^2}
\left[ {\cal W}^{\alpha}(\vec x),\left[{\cal W}_{\alpha}(\vec x),
  \frac{\delta W(C,\vec x)}{\delta {\Phi_i(\vec x)}_{IJ}}
  \right]\right]_{IJ}\right|_{\theta =\bar\theta =0} \right> \, .
\end{align}
Due to the anomaly on the right hand side of (\ref{sevenanom}),
this is not quite an equation in loop space,  although it does
closely resemble the loop equation for a three-dimensional
Chern-Simons theory,  which very formally follows from:
\begin{align}\label{chernloop}
0 = \int DA \frac{\delta}{\delta A^a_i(\vec x)}\left( g^a_{i,\vec
x}[A] e^{-S_{CS}}\right)\, .
\end{align}
If $g_{i,\vec x} = W(C,\vec x)$ is chosen to be a Wilson loop with
a marked point,  (\ref{chernloop}) becomes:
\begin{align}
\frac{k}{4\pi}\left<  {\rm tr}  W(C,\vec x) \epsilon_{ijk}
F_{jk}(\vec x) \right> =
\left< \left. \sum_{I,J}%i\frac{g^2}{32\pi^2}
\frac{\delta W(C,\vec x)}{\delta {\Phi_i(\vec
x)}_{IJ}}\right|_{IJ} \right>\, .
\end{align}
Regularizing and making sense of such equations to obtain the so
called Skein relations is non-trivial \cite{Gambini:1996mb}.

The fact that an anomaly in the loop equations appears only in
7-dimensions, for which our construction involves ${\cal N} =1,
d=4$ superspace,  is consistent with recent conjectures of
Dijkgraff and Vafa. A generalized Konishi anomaly is known to be
crucial for the gauge theory derivation \cite{Cachazo:2002ry} of
the Dijkgraff-Vafa proposal
\cite{Dijkgraaf:2002dh,Dijkgraaf:2002fc} relating the holomorphic
data of ${\cal N} =1, D=4$ gauge theories to large $N$ bosonic
matrix models.  This proposal was extended in
\cite{Dijkgraaf:2003xk} to the case of $D=4+n$ dimensional
theories preserving (at least) ${\cal N} =1, d=4$ supersymmetry.
In this case the holomorphic data was conjectured to be captured
by an auxiliary $n$ dimensional bosonic gauge theory. The action
of this auxiliary gauge theory corresponds to the superpotential
of the $4+n$ dimensional theory written in four dimensional ${\cal
N}=1$ superspace. For the case of the supersymmetric gauge theory
describing a D6-brane wrapping a three-cycle of a Calabi-Yau,  the
auxiliary bosonic gauge theory is Chern-Simons theory on the
three-cycle. Proving the validity of the proposal of
\cite{Dijkgraaf:2003xk} by field theoretic methods (as in
\cite{Cachazo:2002ry}) would presumably require a non-trivial
Konishi anomaly,  which we have found above. It would be very
interesting if one could find a generalization of the chiral
Wilson loops for which the loop equations more closely resemble
those of the large $N$ Chern-Simons theory.

There is a possibly important subtlety in the above discussion,
which is that (\ref{trans}) includes a variation of the gauge
fields.  In practice, one must fix the gauge via a Faddeev-Popov
procedure to define the functional measure.  This amounts to
introducing ghosts and extra terms in the action.  We have
computed the variation of the functional measure without gauge
fixing so that our result appears somewhat formal. However we have
also chosen $g_{i,\vec x}$ so as to get equations involving gauge
invariant operators.  The extra terms in the action do not effect
the classical equations for these operators.  Furthermore the
anomalous term arises from the variation of a part of the
functional measure which does not involve ghosts.  Thus we expect
that the anomaly equation is insensitive to the gauge fixing. .

\section{Conclusions and Remarks}

We have obtained a non-renormalization theorem for a class of
Wilson loops in maximally supersymmetric Yang-Mills in dimensions
$D=3,4,5$ and $6$.  We have made use of the fact that these Wilson
loops belong to the chiral ring associated with a D-3 dimensional
sub-algebra of the full supersymmetry. A non-renormalization
theorem for these loops was conjectured previously by Zarembo
\cite{Zarembo:2002an} in the case of a planar path. We find a
stronger result which includes chiral loops in $\mathbb{R}^3$. It
would be very interesting to understand the stronger result from
the AdS point of view,  where it has not been shown in general
that chiral loops are boundaries of minimal surfaces with zero
regularized area.  In fact this has only been demonstrated for
circular and infinitely long rectangular loops.

The results of this article persist upon taking the three
world-volume dimensions transverse to the d=D-3 dimensional
superspace to be a curved space (such as a special Lagrangian
inside a Calabi-Yau three-fold). Generically, this leaves a theory
preserving only four supercharges in D-3 dimensions. Our results
are unchanged due to the diffeomorphism invariance of the
Chern-Simons superpotential. The non-diffeomorphism invariant
parts of the action contribute terms of the form $\bar D^2
(\cdots)$ in the equations of motion for chiral loops.
Supersymmetry implies that such terms vanish upon taking the
expectation value.

We emphasize that in various instances,  gauge fields may be
manifestly included in chiral operators by making use of a lower
dimensional superspace. This may have interesting applications
besides the one which we have presented.

\section{Acknowledgements}

The authors wish to thank  L. Alvarez-Gaum\'{e}, J. Erdmenger, I.
Kirsch, S. Kovacs, D. L\"{u}st, N. Prezas and K. Zarembo for
enlightening discussions. The work of Z.G. is funded by the DFG
(Deutsche Forschungsgemeinschaft) within the Emmy Noether
programme, grant ER301/1-2. The work of BK is supported by
German-Israeli-Foundation, GIF grant I-645-130.14/1999

%\newpage

\end{document}